# Polarized light from black hole can be a symbol of Cherenkov radiation generated by Faster Than Light movement in vacuum under gravity

Sterling-Jilin Liu, An independent researcher, Email: liujilin@zju.edu.cn, Cell phone: +86 18621356287, Profile: https://www.linkedin.com/in/sterling-liu-9016096b/

**Abstract**
Light speed is assumed as a constant in Einstein's general relativity theory. In this paper, the polarized light from black hole Cygnus X-1 is found fit well with Cherenkov radiation generated by Faster Than Light movement in vacuum under gravity. Thus, proved the gravitational lens is a real existence, the refractive index of space did increase with the gravity field, and light speed is changeable with gravitational potential energy. Possible experiments to double verify the theory on the earth is discussed. A new way to find more advanced alien civilizations in lower gravitational potential energy area is proposed.

**Introduction**

In Einstein's general relativity theory, Mercury precession, gravitational lens and gravitational red shift, etc. can be considered as geodetic effect while assuming light speed is a constant, hence get time goes slower in gravitational field. However, in another view, all these phenomena can be also explained as space curved in gravitational field, refractive index of vacuum is larger, and light speed become smaller, but time keep no change. These two views are both acceptable based on previous experimental verification of general relativity theory, whether light or time speed changed is indistinguishable. In this paper, the author found polarized light from black hole Cygnus X-1 can be well explained by Cherenkov radiation generated by Faster Than Light movement in vacuum under gravity. Thus, proved the gravitational lens is a real existence, the refractive index of space did increase with the gravity field, and light speed is changeable with gravitational potential energy.

**Theory Verification**

P.Laurent reported strongly polarized light from black hole Cygnus X-1, he thought it should be produced by synchrotron radiation, a signature of a strong magnetic field generated by black hole [1]. However, synchrotron assumption from strong magnetic is not necessary, the polarized light can be a symbol of Cherenkov radiation.

Cherenkov found blue glow from an underwater nuclear reactor, which can be explained as electromagnetic radiation emitted when a charged particle passes through a dielectric medium at a speed greater than phase velocity of light in that medium [2]. The light will emit in a cone surrounding particle moving path, the emission angle can be expressed as below, where $n$ is the refractive index of medium, $v$ is particle moving speed, $c$ is speed of light in vacuum.

$$\sin\theta = \frac{c}{nv}$$

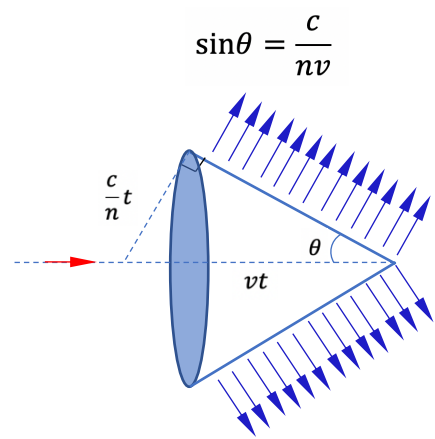

Fig.1 Cherenkov radiation emit in a cone surround particle moving path

The frequency spectrum of Cherenkov radiation by a particle is given by the Frank-Tamm formula [2], The Frank-Tamm formula describes the amount of energy $W$ emitted from Cherenkov radiation, per unit length traveled x and per frequency ω. μ(ω) is the permeability, n(ω) is the index of refraction of the material the charge particle moves through, q is the electric charge of the particle, v is the speed of the particle, and c is the speed of light in vacuum

$$\frac{d^2W}{dx} = \frac{q^2}{4\pi}\mu(\omega)\omega\left(1 - \frac{c^2}{v^2 n^2(\omega)}\right)d\omega$$

According to the Frank-Tamm formula, the emitting power of Cherenkov radiation is increased by light frequency. In general materials, the refractive index n of X-ray or Gamma ray becomes less than 1 and hence no X-ray or Gamma ray emission would be observed [2].

However, based on gravitational lens effect, the refractive index n of vacuum space near massive objects can be viewed as larger than 1 and increased by gravitational field strength. When a charged particle moving with speed close to light, moves from a space with lower gravitational strength into higher gravitational strength area, the velocity of particle can pass through the speed of light in higher gravitational strength area, and Cherenkov radiation will be generated. Different from general materials, the vacuum space refractive index n and μ is the same to all wave lengths in gravitational lens [3], the Frank-Tamm formula can be written as

$$\frac{d^2W}{dx} = \frac{q^2}{4\pi}\mu\omega\left(1 - \frac{c^2}{v^2 n^2}\right)d\omega$$

Define N(ω) as number of protons generated by Cherenkov radiation, considering single proton energy is 2πhω, total N protons radiation power is

$$W = \int_0^\infty 2\pi h\omega N(\omega)d\omega$$

We get

$$\frac{dN(\omega)}{dx} = \frac{q^2\mu}{8\pi^2 h}\left(1 - \frac{c^2}{v^2 n^2}\right)$$

Which means the number of protons N(ω) generated per travel length x is independent of ω, number of protons is the same for different energy gamma rays.

According to the spectrum chart reported by P.Laurent, etc [4] as below left one, the spectrum consists of two components: a "Comptonisation" spectrum caused by photons upscattered by Compton scattering off thermally distributed electrons in a hot plasma (dashed line), and an higher energy component (dash dot line with power law of index 1.6+/-0.2) whose origin is not known [4]. Based on the data, we can get below equation of protons N'(E) across distribution of gamma ray energy E

$$\frac{dN'(E)}{dE} \propto E^{-1.6\pm 0.2}$$

Which can be simplified as

$$N'(E) \propto E^{-0.6\pm 0.2}$$

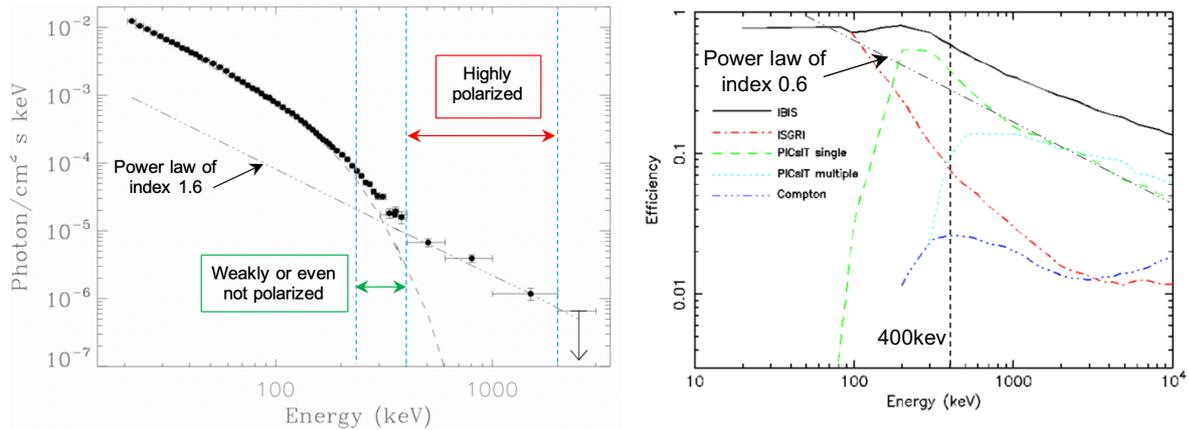

Fig.2 Left one is spectrum received by integral satellite PICsIT detector using histogram mode, integrating gramma ray by time and different energy (keV) channels, and nominalized data is performed on the chart. Right one is efficiency chart of integral satellite gamma ray detector.

The dash dot line gamma distribution with power law of index 1.6+/-0.2 can be well explained as gamma rays generated by Cherenkov radiation. The integral satellite use CdTe and CsI scintillation crystal gamma ray detectors [5]. According to IBIS Observer's Manual of integral satellite [5], the left chart data is got by PICsIT detector using histogram mode, integrating gramma ray by time and different energy (keV) channels, and nominalized data is performed. The efficiency of PICsIT detector follows distribution of above right chart, approximately a power law of index 0.6 (when energy >400keV) because of the nature of scintillation crystal [6][7], which can be written as

$$N''(E) \propto E^{-0.6}$$

We can see *N''(E)* just have the same power law distribution as *N'(E)*, which further proved the gamma ray proton number *N* distribution from balck hole Cygnus X-1 is flat to different gamma ray energy, the signal finally become a power law of index 0.6 distribution is caused by responsing efficiency difference of PICsIT detector to different energy gamma rays.

P.Laurent, etc checked polarization of gamma ray and found the gamma ray band between 250 and 400 keV is weakly or even not polarized, and the band between 400kev and 2000kev is highly polarized, noted in above left chart. This can be also explained as polarized Cherenkov radiation, the polarized gamma ray are immersed by Compton radiation when energy is lower than 400keV, so the measured band 250 ~ 400 keV is weakly polarized, but when proton energy is larger than cut off energy of Compton radiation 400keV, it is high polarized as the nature of Cherenkov radiation.

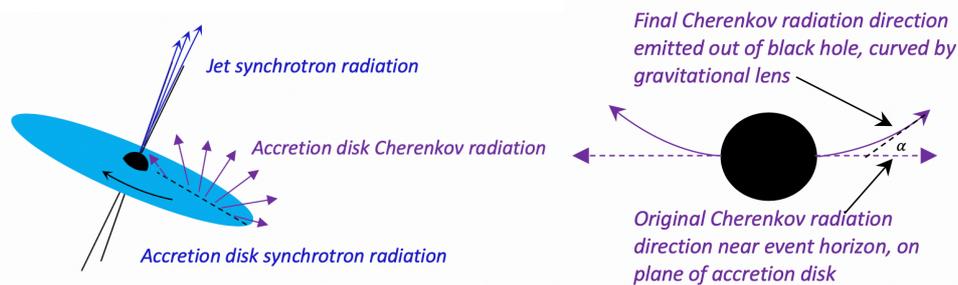

Fig.3 Left one is Cherenkov and synchrotron radiation directions. Right one is Cherenkov radiation curved by gravitational lens effect

P.Laurent, etc observed position angle of the polarized gamma ray electric vector, which gives the direction of the electric field lines projected onto the sky, is 140 ± 15°, has at least 100° deviation away from the previously synchrotron radiation assumption of compact radio jet (position angle of 21~24°) [4]

The position angle can also be well explained with Cherenkov radiation. Considering particles are moving on the plane of accretion disk, as Cherenkov radiation emit light in a cone surrounding particle moving path, the polarized direction is perpendicular to cone surface, so the resultant polarized direction of light is just particle moving direction, have a 90° intersection angle with black hole rotation axis (also the compact radio jet direction). Further considering gravitational lens effect, the original Cherenkov radiation polarized direction will be curved by angle $\alpha$ [8]

$$\alpha = \frac{4GM}{Rc^2}$$

Taking *M* as black hole mass, 14.8 times of sun, 2.94x $10^{31}$kg, *R* as event horizon radius 300km [9], we get α as 16.6°, so the position angle of polarized gamma ray electric vector generated by Cherenkov radiation is 21~24+90+16.6=129.1°, which meet with observed result 140 ± 15°.

**Conclusion and Discussion**

From above discussion, we can conclude that the Cherenkov radiation can fit well with integral satellite observed results in spectrum distribution, polarization and position angle. However, Einstein's general relativity theory defined light speed is a constant to all vacuum space, it is conflict to the fact that Cherenkov radiation is generated by charged particle's Fast Than Light Movement. Thus, we can further derive the gravitational lens is a real existence, the refractive index of vacuum space did increase with the gravity field, so particle moves from outer area to inner area of black hole, the speed will exist light speed of inner area, and hence generate Cherenkov radiation. It means the speed of light varies with different gravitational field strength. Considering gravitational field strength has direction, but light speed is a scalar value, it is better to define light speed changes with gravitational potential energy.

Sun's gravitational potential energy has the maximum variation as 3.3% between winter solstice and summer solstice on the earth. In the past, we did not find significant light difference may because of two reasons. For the first reason, the light speed change may be subtle versus gravitational potential energy change, so a quantitative mode between light speed and gravitational potential energy need to be established in the future work. Maybe near sun surface light speed change will be large enough for measurement. Another reason is the measurement device dimension may decrease the same percentage as light speed in gravitational field, so no difference for measurement result.

In order to further verify the speed of light changed by gravitational potential energy, we can measure received gamma intensity change of a binary system, like the black hole Cygnus X-1 and star HDE 226868, cycle time 5.6days, the gamma ray intensity will also change periodically. The orbital eccentricity is 0.0018, so the intensity change can be 0.18% if assume the light speed is linear proportional to gravitational potential energy. A binary system with higher orbital eccentricity will make the difference detection much easier.

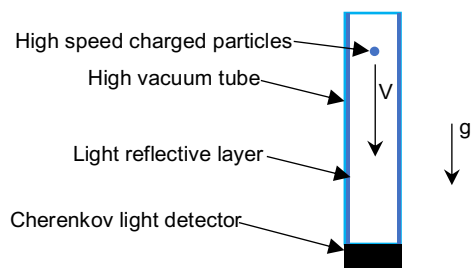

Fig.4 Experiment to verify refractive index changed by gravitational potential energy, detecting Cherenkov radiation of charged particles moving along direction of gravitational force, the vacuum should be as long as possible, in order to get larger gravitational force different and longer particle moving path to generate more radiation

We can also design an experiment as above, apply charged particles moving along direction of gravitational force in a high vacuum tube coated with light reflective layer. The particles can be generated by accelerator or from outer space of earth, move in a speed close to light speed, Cherenkov light will be generated because of gravitational potential energy difference between top and bottom side of tube.

**Prospect**

The earth background gamma radiation may also generated by high speed charged particles from other outer space where gravitational potential energy is lower than earth and solar system, thus the particle speed can be faster than the light speed of earth, generating Cherenkov radiation.

Based on above discussion, we can further infer the more advanced alien civilization than human being will live in a space with lower gravitational potential energy than earth and solar system, as light speed and chemical reaction is much faster, civilization will iterate much faster and much easier for space travel. The probability will be much higher if we search alien civilization in this way, such as near galaxy edge area.